\documentclass{article}

\usepackage[english]{babel}
\usepackage[utf8x]{inputenc}
\usepackage[T1]{fontenc}

\usepackage[a4paper,top=3cm,bottom=2cm,left=3cm,right=3cm,marginparwidth=1.75cm]{geometry}

\usepackage{amsmath}
\usepackage{amssymb}
\usepackage{amsthm}
\usepackage{graphicx}
\usepackage[colorinlistoftodos]{todonotes}
\usepackage[colorlinks=true, allcolors=blue]{hyperref}

\title{Improving RNA secondary structure prediction via state inference with deep recurrent neural networks}
\author{Devin Willmott \and David Murrugarra \and Qiang Ye}
\date{Feb 2020}

\begin{document}

\maketitle
{\footnotesize
     \centerline{Department of Mathematics,
      University of Kentucky,}
  \centerline{Lexington, KY 40506-0027 USA.}
}
\begin{abstract}
The problem of determining which nucleotides of an RNA sequence are paired or unpaired in the secondary structure of an RNA, which we call RNA state inference, can be studied by different machine learning techniques. Successful state inference of RNA sequences can be used to generate auxiliary information for data-directed RNA secondary structure prediction. Typical tools for state inference, such as hidden Markov models, exhibit poor performance in RNA state inference, owing in part to their inability to recognize nonlocal dependencies. Bidirectional long short-term memory (LSTM) neural networks have emerged as a powerful tool that can model global nonlinear sequence dependencies and have achieved state-of-the-art performances on many different classification problems.

This paper presents a practical approach to RNA secondary structure inference centered around a deep learning method for state inference. State predictions from a deep bidirectional LSTM are used to generate synthetic SHAPE data that can be incorporated into RNA secondary structure prediction via the Nearest Neighbor Thermodynamic Model (NNTM). 
This method  produces predicted secondary structures for a diverse test set of 16S ribosomal RNA that are, on average, 25 percentage points more accurate than undirected MFE structures. 
Accuracy is highly dependent on the success of our state inference method, and investigating the global features of our state predictions reveals that accuracy of both our state inference and structure inference methods are highly dependent on the similarity of pairing patterns of the sequence to the training dataset. Availability of a large training dataset is critical to the success of this approach. Code available at \href{https://github.com/dwillmott/rna-state-inf}{https://github.com/dwillmott/rna-state-inf}.

\end{abstract}
\section{Introduction}
The secondary structure of an RNA sequence plays an important role in determining its function \cite{Gardner:2004sf,Mathews:2006uo,Lai:2018cs}, but directly observing RNA secondary structure is costly and difficult~\cite{Chen:2016eu,CBIC:CBIC200300700}. Therefore, researchers have used computational tools to predict the secondary structure of RNAs. One of the most popular methods is the Nearest Neighbor Thermodynamic Model (NNTM)~\cite{Turner:2010uq}. Alternatively, comparative sequence analysis methods~\cite{Gutell:2002kx} use a set of homologous sequences to infer a secondary structure~\cite{Cannone2002}. This method remains the gold standard for secondary structure prediction~\cite{Sukosd:2013rm}.

NNTM is based on thermodynamic optimization to find the secondary structure with the minimum free energy (MFE). There are several implementations of NNTM; some of the popular ones include RNAStructure~\cite{Reuter:2010uq}, GTfold~\cite{Swenson:2012fk}, UNAfold~\cite{Markham:2008fv}, and ViennaRNA package~\cite{Lorenz:2011aa}. However, NNTM has been shown to be ill-conditioned~\cite{Layton:2005fk,Le:1993jt,rogers2017conditioning}. That is, for a given sequence, significantly different secondary structures might exhibit very similar energies. Additionally, the range of accuracies of the predictions of NNTM shows significant variance~\cite{Swenson:2012fk}.

More recently, high-throughput data that correlates with the state of a nucleotide being paired or unpaired has been developed. This data, called SHAPE~\cite{Wilkinson:2008fk} for `selective 2'-hydroxyl acylation analyzed by primer extension', has been incorporated as auxiliary information into the objective function of NNTM with the goal of improving the accuracy of the predictions. This type of prediction is referred to as SHAPE-directed RNA secondary structure modeling~\cite{Deigan:2009fk,Washietl:2012lr}.
The addition of auxiliary information usually improves the accuracy of the predictions of NNTM~\cite{Deigan:2009fk}
but it has been shown that the improvements are correlated with the MFE accuracy~\cite{Sukosd:2013rm}.
The latter result has been obtained by statistical modeling of SHAPE. 
The model in~\cite{Sukosd:2013rm} gives distributions for the values of SHAPE if the state of the nucleotide (as paired or unpaired or helix-end) is known. Thus the model in~\cite{Sukosd:2013rm} can be used to generate SHAPE data for an RNA sequence in silico, given the state of each of the sequence's nucleotides.

In this paper, we present a method for improving the RNA secondary structure prediction based on state inference results. To do so, we first study the problem of determining the state of each nucleotide of an RNA sequence, which we refer to as state inference. State inference is a binary classification task on each nucleotide, which we note is in contrast to full secondary structure inference, which seeks to identify sets of base pairs. We have developed, trained, and tested a deep recurrent neural network that performs this task: given an RNA sequence, the machine outputs a probability that each nucleotide is paired. We can threshold this probability to obtain binary predictions for the state of each nucleotide. 

Additionally, we use the probabilities from the state inference method to generate synthetic SHAPE. Then we use this SHAPE data for directed predictions via NNTM, leading to significant improvements in secondary structure accuracy on sequences where the state inference performed well. We note that our approach for generating SHAPE is different from the statistical models in~\cite{Sukosd:2013rm}, which generate synthetic SHAPE data by sampling from the distribution models.

We note that other deep learning methods for the problem of RNA secondary structure inference have been explored \cite{willmottthesis}. Although we are primarily interested in using state inference to direct secondary structure predictions, there exist other motivations for state inference. For example, such a method could be used to identify binding sites in RNA-RNA interactions~\cite{Tafer:2011aa,dichiacchio2015accessfold}.

\begin{figure}[ht]
\begin{center}
\includegraphics[width=5.5in]{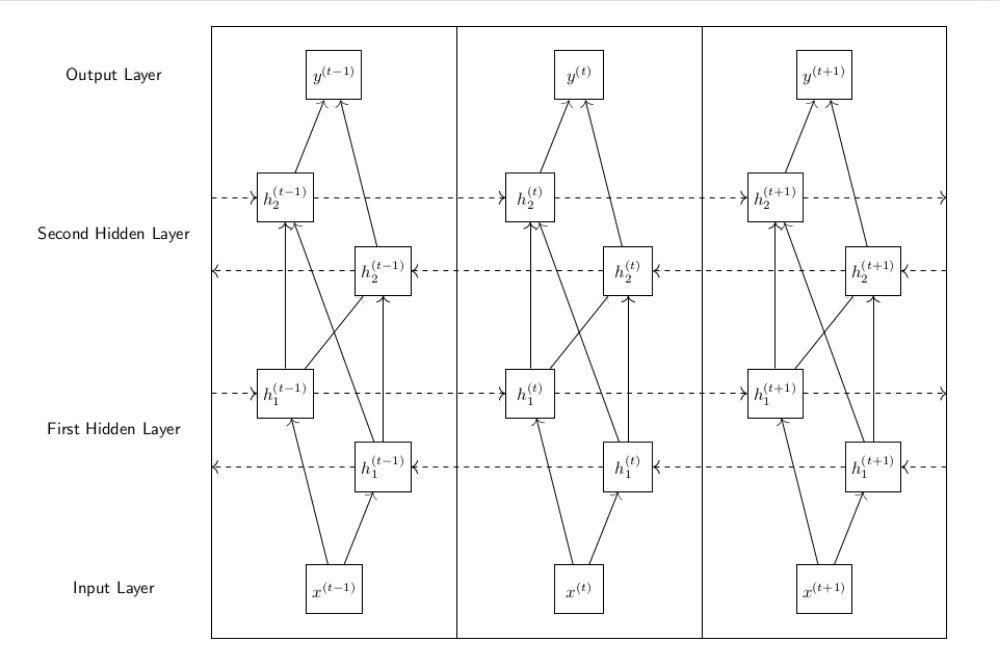}    
\end{center}
\caption{Bidirectional RNN.
Diagram of a bidirectional RNN with two hidden layers for state inference at three different timesteps. Here $x$ denotes the input, $h_1$ the first layer hidden variable, $h_2$ the second layer hidden variable, and $y$ the output, with superscripts representing timesteps and arrows showing the propagation of information through the network. Each hidden layer combines information from the previous layer, earlier and later timesteps, and an internal memory state to compute its output.}
\end{figure}

\section{Methods}

\subsection{SHAPE-Directed NNTM}

Under the nearest neighbor thermodynamic model (NNTM), stacks and loops are each assigned a free energy based on base pair type (Watson-Crick or wobble), with pairs contributing negative energy, and loops contributing positive energy. The energy of a secondary structure is the sum of the energies of these local features. Dynamic programming methods can be used to efficiently find the secondary structure with the smallest energy, called the minimum free energy (MFE) structure, which is usually taken as the predicted structure by the NNTM method.

The SHAPE data comes from high-throughput chemical probing experiments and associate a reactivity value to each nucleotide of an RNA sequence. It has been found experimentally that high SHAPE values are correlated with unpaired positions, and small values with paired position. SHAPE has been incorporated into NNTM by adding a pseudo-free energy term to the model~\cite{Deigan:2009fk}. This term is generated by the following relation:
\begin{equation}
\label{eq:shape_term}
\Delta G_{\text{SHAPE}}(i) = 2.6\cdot\ln(\text{SHAPE}(i)+1) - 0.8 
\end{equation} 

This energy is added to base pair stacks involving nucleotide $i$. In effect, the term $\Delta G_{\text{SHAPE}}(i)$ penalizes base pairs involving nucleotides with high SHAPE values, and encourages base pairs among nucleotides with low SHAPE values. Incorporating SHAPE data consistently leads to significant increases in the accuracy of the MFE structure~\cite{Deigan:2009fk,Washietl:2012lr,Sukosd:2013rm}.

\subsection{Directing NNTM With State Information}

We are interested in using deep learning tools in tandem with SHAPE direction to improve secondary structure inference. However, deep learning methods require extraordinary large datasets; the relative dearth of available experimental SHAPE data prohibits us from directly learning synthetic SHAPE data with a neural network. Instead, our proposed method for secondary structure inference uses a viable method for state inference with the ability to influence the NNTM energy function on a per-nucleotide basis via SHAPE direction. 
Note that the same limitation applies to the secondary structure data for which a large dataset is available for 16S sequences only but not for other RNA types; see Section 4.4 for more details. 

Our method is a three-step process: 

\begin{enumerate}
\item A machine learning method for predicting the state of each nucleotide in a sequence;
\item A function converting these state predictions into artificial SHAPE data;
\item The SHAPE-directed NNTM function that takes both the original RNA sequence and the generated SHAPE data and outputs a predicted secondary structure. 
\end{enumerate}

Note that Step 3 is a well established method and the novelty here is in deriving the artificial SHAPE data in Steps 1 and 2. 
For the task of state inference, we trained a deep neural network using a set of known RNA sequences and structures that generates a sequence of state predictions, detailed in Section 2.3. The output of this neural network is a sequence $p$ of the same length as the original RNA sequence, where $p(i)$ is the predicted probability that the nucleotide in position $i$ is paired.

With these predictions in hand, we convert each predicted probability $p(i)$ to a SHAPE value to be associated with nucleotide $i$. To construct a function for this purpose, we note that a SHAPE value of $\approx 0.3603$ will not contribute any positive or negative energy to the NNTM energy function; this can be seen by setting $\Delta G_{\text{SHAPE}}(i)$ to $0$ in Equation~\ref{eq:shape_term} and solving for $\text{SHAPE}(i)$~\cite{Eddy:2014fj}. We would therefore like to assign predictions of $0.5$ to a SHAPE value of $0.3603$, as these predictions give no information as to the state of the nucleotide. With this in mind, we use the following piecewise linear function to generate SHAPE, where $a$ and $b$ are constants to be specified.

\begin{equation}
\label{eq:shapegeneration}
f(i) = \begin{cases} 2(0.3603-b)p(i) + b, & {\rm if}\; \; 0 < p(i) \leq 0.5 \\ 2(a-0.3603)(p(i) - 1) + a, & {\rm if}\; \; 0.5 < p(i) \leq 1 \end{cases}
\end{equation}

This function has range $[a,b]$, with $f(i) = a$ if $p(i)=1$, $f(i) = b$ if $p(i)=0$, and $f(i) = 0.3603$ if $p(i) = 0.5$. To determine values of $a$ and $b$, we considered experimentally collected SHAPE data from two E. coli sequences, one 16S sequence and one 23S sequence~\cite{Sukosd:2013rm}. Together, these two sequences contain a a total of 4187 nucleotides, and represent a wide variety of structural motifs. We took the mean SHAPE value among both paired nucleotides and unpaired nucleotides; these values are $0.214$ and $0.6624$, respectively.

All of our experimental results in Section 3 will use $a = 0.214$ and $b = 0.6624$ in our SHAPE generation function. These choices are motivated by real SHAPE values, and thus are sensible estimations of the best values. However, they may not be the optimal values for our purposes. In Section 4.2, we explore how varying these values may affect the accuracy of our predicted secondary structures. These experiments indicate that the $a$ and $b$ values used in our experimental results are nearly optimal.

With set values of $a$ and $b$, we can generate a sequence of artificial SHAPE data. We then use SHAPE-directed NNTM as described in the previous section to obtain our secondary structure prediction.

\subsection{State Inference with Deep Neural Networks}

Neural networks are tools from the realm of machine learning for solving classification and regression problems. In a neural network, model parameters are trained using a dataset of known input-output pairs: we define a loss function based on the difference between machine predictions and target outputs, retrieve gradient directions for parameters with respect to this loss using the backpropagation algorithm~\cite{rumelhart1988learning}, and optimize parameters using iterative first order methods, such as gradient descent.

Recent advances in machine learning come primarily from deep neural networks~\cite{Goodfellow-et-al-2016}, which are stacks of multiple neural networks: the output of one neural network in the stack acts as the input for the next. Each of these constituent neural networks is referred to as a layer of the deep neural network. These multiple layers allow the deep neural network to learn and represent complex nonlinear relationships among inputs.

For the task of state inference, we use a deep recurrent neural network (deep RNN). RNNs work specifically with sequential data by combining the learning methods of neural networks with the architecture of a discrete-time dynamical system. A single layer RNN has a state $h^{(i)}$ that is a function of the state at previous time step $h^{(i-1)}$ and the input  $x^{(i)}$ at time $i$. The function is a composition of an elementwise nonlinear activation function with a linear map. Sequence elements are fed to the machine as inputs one at a time; at time $i$, the machine receives the nucleotide in position $i$, given as a one-hot encoding $x^{(i)}$. From this input and the previous state $h^{(i-1)}$,  machine parameters generate a state $h^{(i)}$ that encodes the pairing information up to step $i$, from which another function  produces 
the output $y^{(i)}$, the machine prediction of the probability that nucleotide $i$ is paired.

We make a number of modifications to our deep RNN to increase state inference accuracy. Most notably, we use a popular variant of RNN called the Long Short-Term Memory (LSTM) architecture~\cite{Hochreiter:1997:LSM:1246443.1246450}, which incorporates a gating mechanism and a memory cell to increase accuracy beyond that of traditional RNNs on a variety of sequential learning tasks. Specifically, the LSTM gating architecture allows a certain component of the state to directly pass into future steps, maintaining the flow of state information or memory over the long term. This significantly increases machine capability to model long term dependence and hence capture potential long-range base pairs~\cite{Li:2018qf}.  We also make this network bidirectional~\cite{schuster1997bidirectional,graves2005framewise}; this is a minor modification that allows information to flow both forward and backward through the sequence. A two-layer bidirectional RNN is shown in Figure 1.

\subsection{Dataset, Implementation, and Metrics}

Our experiments will focus on a test set of sixteen 16S ribosomal RNA sequences used in SHAPE direction experiments in~\cite{Sukosd:2013rm}. Sequences in this set have a wide range of NNTM accuracies.

Our deep neural network requires a large dataset of RNA sequences with known states from which to learn. For this task, we used secondary structure data from the Comparative RNA Web site, run by the Gutell Lab at the University of Texas~\cite{cannone2002comparative}. This site hosts a collection of known RNA sequences and secondary structures obtained using comparative sequence analysis. Compiling all of the available 16S rRNA results in a set of 17032 sequences and a total of over 21 million nucleotides. We refer to this as the CRW dataset.

To ensure that our model does not simply memorize large portions of sequences in the test set, we compared each CRW dataset sequence with each test set sequence and removed CRW sequences with significant similarities prior to training. In this filtering process, if the two sequences have a common block of nucleotides of more than 10\% of the length of the test sequence, or if the two sequences can be aligned such that they have common nucleotides accounting for more than 80\% of nucleotides of the shorter sequence, we remove it from the training set. See available code for additional details. This process leaves us with 13118 sequences and a total of approximately 16.5 million nucleotides, with a mean and median sequence length of 1264 and 1431, respectively. We then split this set into two random halves to produce a training and validation set.

For the sake of comparison, we also trained and tested a number of higher-order hidden Markov models (HMM) using the same training, validation, and test sets used by the neural network. In a $k$ order HMM, hidden state transitions depend on the previous $k$ states in the sequences, rather than only the previous state; this improves their representational capacity, but at the cost of a model size that increases exponentially with $k$. We trained our HMMs with maximum likelihood estimation over the training set, and performed inference with the Viterbi algorithm~\cite{durbin1998biological}, a backtracking algorithm that exploits the Markovian nature of HMMs to efficiently produce the likeliest state sequence under the model's transition probabilities. HMMs are fundamentally incapable of recognizing dependencies across many timesteps, and we therefore expect the deep neural network to outperform the HMM. However, they provide a baseline against which to measure neural network output.
We note that the HMM formalism has been used before for state inference of RNA but for a different purpose~\cite{Selega:2017ty,Ledda_2018,Kawaguchi:2019qy}. Code of the HMM implementation is available at \href{https://github.com/dwillmott/rna-state-inf}{https://github.com/dwillmott/rna-state-inf}.

We implemented a variety of deep recurrent networks in Keras~\cite{chollet2015keras}, a Python deep learning API, with Theano~\cite{2016arXiv160502688short} as a backend; code is available at \href{https://github.com/dwillmott/rna-state-inf}{https://github.com/dwillmott/rna-state-inf}. We found a four-layer network to be the optimal balance of representational capacity and training speed. The largest layers are the middle two, which are both bidirectional LSTMs. The first and last layers are small one-dimensional convolutional layers~\cite{lecun1998gradient,Goodfellow-et-al-2016}. These layers act as learnable pre- and post-processing convolutions; they take in and process local information in small regions of the sequence, and allow the two recurrent layers to focus on long-range dependencies across many timesteps. The output dimension of each of the machine's four layers at each timestep are $100$, $400$, $100$, and $2$, respectively, giving a machine with a total of 595,552 trainable parameters, which we trained using binary cross entropy loss and RMSprop~\cite{tieleman2012lecture}, a gradient-descent like training algorithm. See available code material for hyperparameters and training modifications.

Our code works quite efficiently for the kind of sequences tested in this paper. On a 2080 GeForce RTX 2080 Ti GPU, training takes about 4 hours, and once trained, state inference takes about 0.005 seconds for a 5S sequence ($\approx$ 150 nucleotides), 0.05s for a 16S sequence ($\approx$ 1500 nucleotides), and 0.1s for a 23S sequence ($\approx$ 3000 nucleotides), which are roughly linear with the size of the sequence.

With the sequence of state prediction probabilities $p$, we generated artificial SHAPE data using Equation~\ref{eq:shapegeneration}. Finally, we used both the original sequence and the generated SHAPE data as input for GTfold~\cite{Swenson:2012fk}, an efficient NNTM implementation, to compute a SHAPE-directed MFE structure. GTFold is used in this work for all NNTM experiments and results.

When evaluating predicted secondary structures, we compare its set of base pairs with those of the native structure. A predicted base pair is counted as true positive (TP) if it exists in both the predicted and native structure, a false positive (FP) if it appears in the predicted structure but not in the native structure, and a false negative (FN) if it appears in the native structure and not in the predicted structure. We report on three measures of performance: PPV, the proportion of true positives in the predicted structure ($\frac{\text{TP}}{\text{TP} + \text{FP}}$); sensitivity, the fraction of true positives in the native structure ($\frac{\text{TP}}{\text{TP} + \text{FN}}$); and accuracy, the arithmetic mean of PPV and sensitivity: $\frac{1}{2}(\frac{\text{TP}}{\text{TP} + \text{FP}}+\frac{\text{TP}}{\text{TP} + \text{FN}})$ \cite{Gardner:2004sf}.

In later sections, we will also consider the performance of our deep learning methods for state inference. Unlike secondary structure inference, which classifies base pairs, state inference is a binary classification of each nucleotide. When discussing the accuracy of state predictions, we will define accuracy to be the proportion of true predictions among all predictions in the sequence. This is  distinct from notions of secondary structure inference accuracy, and so the metrics on state inference and structure inference cannot be directly compared. 

\section{Results}

\subsection{Native State Directed NNTM}

\begin{table}[!t]
\centering
\caption{Structure Inference Results on the test set\label{Tab:04}}
\setlength{\tabcolsep}{4pt}
\begin{tabular}{|l|c|c|c|c|}
\cline{1-5}
Sequence Name & \multicolumn{1}{c|}{Undirected} & \multicolumn{3}{c|}{Directed MFE} \\
\cline{3-5}
 & MFE & Predicted & S{\"u}k{\"o}sd & Native \\
\cline{1-5}
                   E. cuniculi & 0.171 & 0.183 & 0.273 & 0.336 \\
                   V. necatrix & 0.181 & 0.314 & 0.503 & 0.705 \\
                    C. elegans & 0.203 & 0.248 & 0.308 & 0.519 \\
                   E. nidulans & 0.272 & 0.325 & 0.601 & 0.832 \\
                    N. tabacum & 0.323 & 0.692 & 0.593 & 0.859 \\
                Cryptomonas.sp & 0.339 & 0.838 & 0.739 & 0.898 \\
              Synechococcus.sp & 0.361 & 0.848 & 0.697 & 0.885 \\
                   M. musculus & 0.375 & 0.397 & 0.509 & 0.782 \\
              M. gallisepticum & 0.385 & 0.849 & 0.721 & 0.889 \\
                       E. coli & 0.411 & 0.852 & 0.744 & 0.880 \\
                   B. subtilis & 0.512 & 0.848 & 0.753 & 0.881 \\
              D. desulfuricans & 0.533 & 0.875 & 0.724 & 0.898 \\
                C. reinhardtii & 0.537 & 0.845 & 0.702 & 0.868 \\
                   T. maritima & 0.562 & 0.881 & 0.733 & 0.896 \\
                      T. tenax & 0.619 & 0.766 & 0.754 & 0.861 \\
                   H. volcanii & 0.752 & 0.864 & 0.809 & 0.907 \\
\cline{1-5}
                       Mean    & 0.408 & 0.664 & 0.635 & 0.806 \\
                       Median  & 0.380 & 0.841 & 0.712 & 0.874 \\
\cline{1-5}
\end{tabular}
\begin{flushleft}
Table of accuracy of MFE structures using NNTM with a variety of SHAPE directions. First column: undirected MFE. Second column: predicted state directed MFE (see Section 3.2). Third column: mean performance of sampled SHAPE directed NNTM in~\cite{Sukosd:2013rm}. Fourth column: native state directed NNTM (see section 3.1).
\end{flushleft}
\end{table}

Before analyzing the results of the entire pipeline of our method, we first examined our SHAPE generation function in detail. To do so, we used the native state of each sequence in our test set to generate SHAPE. This was done by setting $p(i)$ to $1$ if the nucleotide in position $i$ is paired, and $p(i)$ to $0$ if it is unpaired. We then use Equation 2 to generate artificial SHAPE. This will result in a generated SHAPE value of $0.6624$ for all paired nucleotides and $0.214$ for unpaired nucleotides, which we then use to direct NNTM. We refer to the resulting predicted structures as native state directed MFE.

This experiment is similar to those run in~\cite{Sukosd:2013rm}, and uses the same set of data to choose appropriate SHAPE values. The difference is in the method of SHAPE generation: whereas that paper constructs SHAPE distributions from the data and stochastically samples from these distributions, we use the mean of paired and unpaired nucleotides' SHAPE values.

The results of this experiment reinforce many of the findings in~\cite{Sukosd:2013rm}. A comparison of accuracy of all three methods (undirected MFE, stochastically directed MFE from~\cite{Sukosd:2013rm}, and native state directed MFE) is available in Table~\ref{Tab:04}. Overall, native state directed MFE structures are highly accurate, with twelve of the sixteen test sequences enjoying accuracy above 80\%. Both direction methods are an improvement on the accuracy of the undirected MFE structure for every test set sequence, and native state directed accuracy represents a further improvement from the stochastic model in~\cite{Sukosd:2013rm}. In the case of native state direction, accuracy improvements over undirected MFE range between 15 percentage points (H. volcanii) and 57 percentage points (Cryptonomas.sp). Consistent with observations in~\cite{Sukosd:2013rm}, greatest increases are concentrated in sequences with middling undirected MFE accuracy; for sequences with undirected accuracy between 25\% and 45\%, native state directed MFE accuracy is an improvement by more than 40 percentage points.

This experiment is equivalent to assuming that our deep learning state inference method has perfect performance, and as such we can interpret the accuracy of native state directed MFE structures to be an upper bound on the performance of our method. On average, the high accuracy exhibited in this experiment gives strong evidence that there are large potential gains in MFE accuracy to be made with our method. However, several sequences with low undirected MFE accuracy sequences like E. cuniculi and C. elegans are known to be particularly resistant to SHAPE direction~\cite{Sukosd:2013rm}, and this is reflected in relatively poor native state directed MFE accuracy. We thus cannot expect our method to exhibit large improvements over undirected MFE structures in these cases.

\begin{table}[b]
\centering
\caption{State inference accuracy of neural network vs. HMM on validation and test sets\label{Tab:validset}}
\setlength{\tabcolsep}{2pt}
\begin{tabular}{|c|c|c|c|c|c|c|}
\hline
 & \multicolumn{3}{c|}{Validation Set} & \multicolumn{3}{c|}{Test Set} \\
\cline{2-7}
Machine & Acc & PPV & Sen & Acc & PPV & Sen \\ 
\hline
Order 1 HMM & 0.623 & 0.632 & 0.852 & 0.612 & 0.646 & 0.767 \\
Order 2 HMM & 0.662 & 0.671 & 0.826 & 0.651 & 0.686 & 0.759 \\
Order 3 HMM & 0.674 & 0.693 & 0.794 & 0.672 & 0.713 & 0.750 \\
Order 4 HMM & 0.685 & 0.714 & 0.771 & 0.684 & 0.729 & 0.742 \\
Order 5 HMM & 0.684 & 0.711 & 0.776 & 0.683 & 0.730 & 0.742 \\
Neural Network & 0.954 & 0.950 & 0.972 & 0.839 & 0.858 & 0.873 \\
\hline
\end{tabular}

\begin{flushleft}
A comparison of accuracy, PPV, and sensitivity of output from our LSTM-based neural network for state inference compared with those of HMMs of various orders. The composition of validation and test sets is described in Section 2.4.
\end{flushleft}
\end{table}

\subsection{Predicted State Directed NNTM}

\begin{table*}[!ht]
\centering
\caption{State inference results on the test set from LSTM vs. HMM \label{Tab:testset}}
\setlength{\tabcolsep}{4pt}
\begin{tabular}{|l|cc|cc|cc|}
\cline{1-7}
 & \multicolumn{2}{c|}{Accuracy} & \multicolumn{2}{c|}{PPV} & \multicolumn{2}{c|}{Sensitivity}\\
Sequence Name & LSTM & HMM & LSTM & HMM & LSTM & HMM \\
\cline{1-7}
                   E. cuniculi & 0.680 & 0.661 & 0.713 & 0.693 & 0.774 & 0.773\\
          Vairimorpha necatrix & 0.661 & 0.600 & 0.721 & 0.689 & 0.683 & 0.576\\
                    C. elegans & 0.558 & 0.584 & 0.570 & 0.613 & 0.624 & 0.552\\
           Emericella nidulans & 0.657 & 0.584 & 0.692 & 0.681 & 0.741 & 0.539\\
             Nicotiana tabacum & 0.913 & 0.705 & 0.917 & 0.734 & 0.938 & 0.787\\
                Cryptomonas.sp & 0.926 & 0.676 & 0.935 & 0.730 & 0.941 & 0.728\\
              Synechococcus.sp & 0.938 & 0.700 & 0.943 & 0.740 & 0.953 & 0.769\\
                   M. musculus & 0.608 & 0.603 & 0.626 & 0.655 & 0.637 & 0.520\\
      Mycoplasma gallisepticum & 0.919 & 0.639 & 0.933 & 0.713 & 0.932 & 0.668\\
                       E. coli & 0.924 & 0.699 & 0.937 & 0.742 & 0.938 & 0.774\\
             Bacillus subtilis & 0.973 & 0.698 & 0.979 & 0.731 & 0.976 & 0.788\\
   Desulfovibrio desulfuricans & 0.926 & 0.712 & 0.940 & 0.741 & 0.938 & 0.803\\
     Chlamydomonas reinhardtii & 0.906 & 0.687 & 0.915 & 0.725 & 0.928 & 0.761\\
           Thermotoga maritima & 0.931 & 0.752 & 0.944 & 0.760 & 0.943 & 0.864\\
           Thermoproteus tenax & 0.818 & 0.782 & 0.845 & 0.785 & 0.866 & 0.894\\
                   H. volcanii & 0.782 & 0.739 & 0.809 & 0.769 & 0.841 & 0.820\\
\cline{1-7}
Average & 0.820 & 0.676 & 0.839 & 0.719 & 0.853 & 0.726\\
  Total & 0.839 & 0.684 & 0.858 & 0.729 & 0.873 & 0.742\\
\cline{1-7}

\end{tabular}
\begin{flushleft}
Table of accuracy, PPV, and sensitivity for our LSTM-based state inference model vs. an order 4 HMM. Sequences are arranged in ascending order of MFE accuracy as an indication of the difficulty of secondary structure inference for each sequence. Average indicates the average metric for each sequence, while Total gives the total metrics for all nucleotides in the test set. 
\end{flushleft}
\end{table*}

We now use the predictions from our deep neural network to generate SHAPE data that will in turn direct NNTM; we refer to these predictions as predicted state directed MFE structures. We emphasize that, unlike the native state direction explored in the previous section, this method does not assume prior knowledge of the state of the sequence, and thus represents a practical method of secondary structure inference.

The results of applying our method to the sequences in the test set are available in Table~\ref{Tab:04}, which indicates that the extraordinary gains from native state directed NNTM are not always preserved in practice. Predicted state directed structures fall into two clear categories: five are quite inaccurate, with accuracy below 40\%, while among remaining eleven structures are all near or above 70\%, and nine of these are above 80\%. Even with the high variance of accuracies among these structures, predicted state directed MFE structures are 25 percentage points more accurate than undirected MFE structures on average, and every sequence in the test set experiences some increase in accuracy. However, these improvements vary greatly, with several sequence staying within 5\% of undirected MFE accuracy, while for four other sequences we improve by more than 40\%, with the highest improvement (Cryptonomas.sp) at 50\%.

To some extent, poor accuracy is explained by our experiment with native state directed NNTM. Indeed, the five sequences with poor accuracy from our method are the five worst-performing with native state directed  using native states to direct NNTM, and only one of these exceeds 80\% with native states. At worst, native state directed NNTM gives only 34\% accuracy for E. cuniculi and 52.6\% for C. elegans, and this ceiling is much lower than accuracy achieved in many of our other predicted structures. Thus, the difficulties of our method with these sequences may be attributed to the RNA sequences themselves that are particularly challenging for structure inference as discussed in the directability of NNTM section in  \cite[p.2812]{Sukosd:2013rm}. However, in these cases and some others (E. nidulans, M. musculus), predicted state directed MFE accuracy does not come close to native state directed MFE accuracy. This is in contrast to our highest performing sequences (D. desulfuricans, T. maritima), where predicted state directed MFE is within several percentage points of native state directed MFE accuracy.

\begin{figure}[ht]
\begin{center}
\includegraphics[width=5.5in]{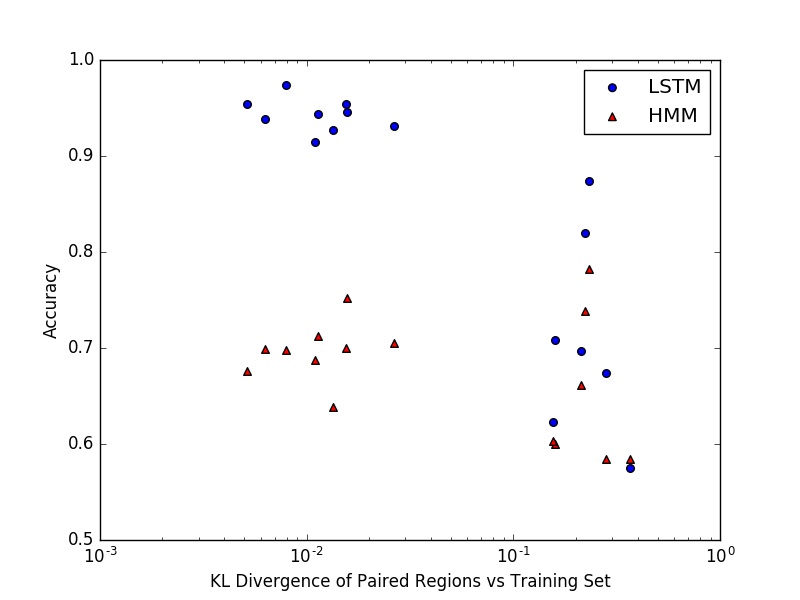}    
\end{center}
\caption{Test Set KL Divergence.
Plot comparing each test set sequence's LSTM (neural network) and HMM state inference accuracy vs its Kullback-Leibler divergence from training set paired region distribution. KL divergence was calculated as KL$(P \| Q)$, where $P$ is the test sequence distribution and $Q$ is the training set distribution.} \label{kl}
\end{figure}


\begin{figure}[h!]
\begin{center}
\includegraphics[width=2.8in]{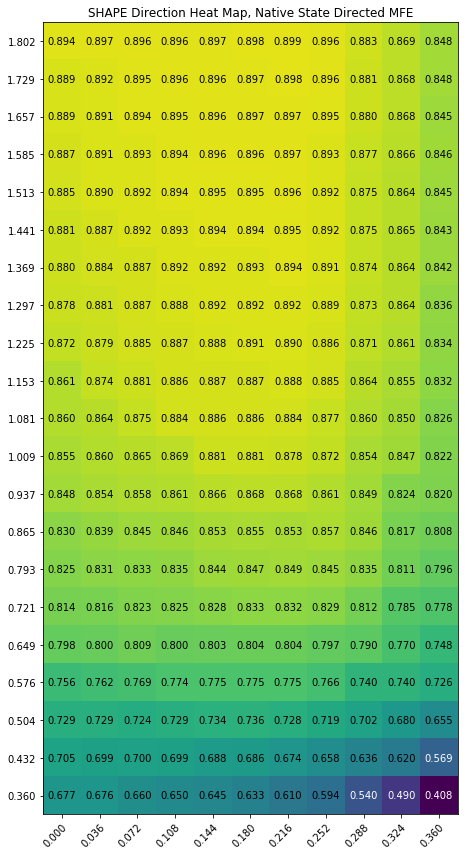} \hspace{0.15in} \includegraphics[width=2.8in]{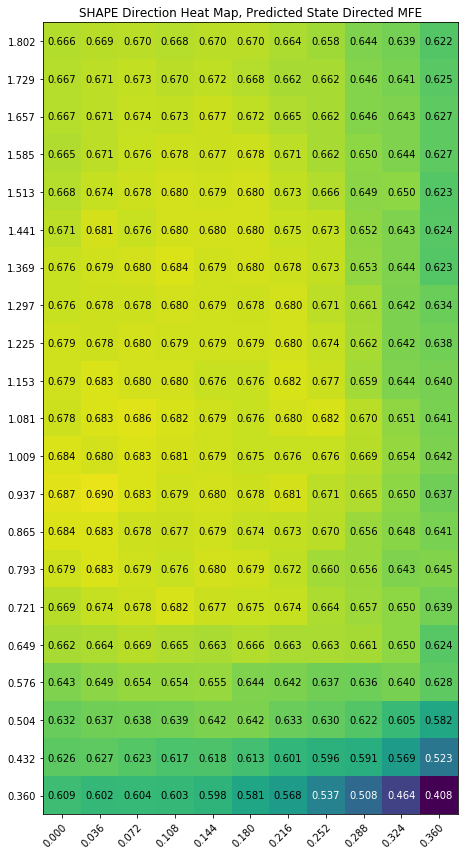}
\end{center}
\caption{SHAPE Direction Heat Maps.
Accuracy of native state directed MFE (left plot) and predicted state directed MFE (right plot) for various ranges $[a,b]$ of output from our SHAPE generation function. In each, the lower right corner corresponds to $a=b=0.3603$, which is equivalent to no SHAPE direction.} \label{heatmaps}
\end{figure}

\section{Discussion}

\subsection{State Inference Accuracy}

The foundation of our method is our deep neural network for state inference: this network provides probabilities that are converted into a pseudo-free energy term in the NNTM energy function. To understand the sources of high and low performance of our structure inference method, we can directly evaluate the output of our deep neural network for state inference, using higher order HMMs as a baseline.


To calculate the accuracy of the neural network's output, we thresholded each prediction $p(i)$ above and below $0.5$, taking $p(i) > 0.5$ to be a positive prediction and $p(i) < 0.5$ to be a negative prediction. The accuracy, PPV, and sensitivity of both neural network and HMM predictions are shown in Table~\ref{Tab:validset}. Though the table exhibits an upward trend in accuracy as the order of the HMM increases, we found that accuracy plateaued and eventually decreased beyond order $5$. As expected, the LSTM clearly outperforms HMMs of all orders on the validation set. More importantly, this is the case for our test set as well, where the LSTM outperforms the best HMM in accuracy by nearly 13\%.

The order $4$ HMM exhibits the highest accuracy on the validation set. We further compared state inference accuracy on each test set sequence using both the order 4 HMM and the neural network. The accuracy, PPV, and sensitivity of these predictions are shown in Table~\ref{Tab:testset}. The accuracy of neural network state predictions were, on average, 15 percentage points higher than that of the HMM, and was higher for every sequence but one (C. elegans). Table~\ref{Tab:testset} orders sequences in ascending order of undirected MFE accuracy; however, this ordering reveals no straightforward relationships among neural network state inference accuracy, HMM state inference accuracy, and MFE structure accuracy. Neural network accuracy varies much more among sequences: the difference between the sequences with lowest and highest accuracy (C. elegans and B. subtilis, respectively) is more than 40 percentage points. Sequences can be grouped according to accuracy: poor (below 70\%) for five sequences, medium (near 80\%) for two more, and high (above 90\%) for the remaining nine.

As expected, state inference accuracy  exhibits a strong effect on predicted state directed MFE accuracy. State inference accuracy above 90\% means that our predicted states are quite close to native states; consequently, predicted state direction and native state direction (equivalent to 100\% accuracy) should produce similar predicted structures in these cases, as evidenced by their difference of only a few percentage points in Table~\ref{Tab:04}. Meanwhile, the five sequences with poor state inference accuracy are exactly those where predicted state directed MFE accuracy is below 40\%.

The effect of state inference accuracy is particularly evident when considering the improvement over undirected MFE accuracy: for four of the five sequences with poor accuracy (all but V. necatrix), predicted state directed MFE accuracy is within 6 percentage points of undirected MFE accuracy. For V. necatrix and both sequences with medium state inference accuracy, predicted state direction improves structure accuracy by 10-15 percentage points. The remaining nine sequences all have high state inference accuracy, and their directed structures are 30 percentage points more accurate than undirected MFE.

We note an interesting relationship between native state directed MFE accuracy and our neural network's state inference accuracy. The five sequences with state inference accuracy below 70\% are the five worst performing sequence when predicting structure with native state directed NNTM. This suggests that there may be fundamental difficulties in understanding pairing structures of these sequences.

\subsection{Paired Regions \& Global Structure}

Our metrics in Table~\ref{Tab:testset} give us an idea of the proportion of correct machine predictions on individual nucleotides' states, but they do not indicate whether predictions produce state sequences that preserve global properties, such as patterns of paired and unpaired states. In particular, we want the number and sizes of paired and unpaired regions of the state sequence prediction to match those in the original. A paired region in the state roughly corresponds to one half of a helix in the secondary structure, so we theorize that recognizing this information is vital for producing state predictions that successfully aid structure inference.

We considered the distribution of sizes of paired regions in each test set state sequence, and compared them to the distributions of neural network and HMM state predictions. Despite larger variance in state inference accuracy, we found evidence that the neural network was, on average, much more capable than the HMM of capturing this global structure. The median size of paired region in neural network predicted state differed from the median in the native state by at most one for every test set sequence, while the HMM's median paired region size was routinely several nucleotides larger. 
The neural network also performs better in predicting the total number of paired regions in the state, producing predictions that, on average, had 6 more paired regions than the native state, while HMM predictions had an average of 57 fewer regions. 

We note that this discrepancy is to be expected in the context of nonlocal interactions. Paired region size is exactly the sort of nonlocal feature that HMMs cannot predict: at a given time, the HMM does not know how long it has been outputting positive predictions, and is thus limited in its capacity to detect large paired regions.

Considering the non-locality of paired regions can help to explain the poor performance of the neural network on certain test set sequences. High neural network accuracy is nearly always accompanied by a particular type of distribution of large paired regions: one of length 17, one of length 13, and several more of length 12 and 11. 
In contrast, this pattern does not hold for those with low or medium state inference accuracy of the remaining seven, all have either paired regions of length larger than 20 
or very few paired regions of length larger than 10.

We can compare the distribution of the lengths of paired regions in each of our test sequences to the distribution in the training set. We find that the training set overwhelmingly contains sequences with paired region distributions similar to the test set sequences on which the neural network performs well. In particular, we note that the training set has relatively few large paired regions: in the entire training set, there are 5 regions of length 18, 2 regions of size 19, 4 regions of size 20, and none larger than 20. Thus, during training the machine is penalized for outputting more than 20 contiguous positive predictions. Consequently, neural network predictions do not create sufficiently large regions for many test set sequences.

To quantify this difference, we considered the Kullback-Leibler (KL) divergence of the distribution of the paired region lengths between the entire training set and the distribution for each test set sequence. The KL divergence measures the similarity of each test set sequence's paired region distribution as compared with the distribution of the entire training set. Figure~\ref{kl} plots state inference accuracy for each machine and test set sequence against its KL divergence.

Two clusters of sequences emerge in this plot: one with KL divergence near $0.01$, and another with KL divergence near $0.5$. All nine sequences with high state inference accuracy are in the former cluster, while the seven low and medium accuracy sequences are in the latter. The disparity in neural network accuracy and HMM accuracy on sequences with more similarity to the training set suggests that the increase in neural network performance comes from its ability to recognize global structure in these sequences. On the other hand, neural network accuracy is only a modest improvement from HMM accuracy in the low to medium accuracy cluster, where global structure diverges significantly from that of the training set. 

\subsection{Modifying Synthetic SHAPE Values}

Our method uses SHAPE-directed NNTM as a means of assigning pseudo-free energies to individual nucleotides. All of our results in Section 3 assign nucleotides a SHAPE value in the range $[0.214, 0.6624]$. These endpoints are based on the mean SHAPE value of paired and unpaired nucleotides from 16S and 23S E. coli sequences. However, our method converts to SHAPE primarily as a means of assigning pseudo-free energies to individual nucleotides with NNTM, and not as a genuine attempt to generate plausible SHAPE data. Thus, the endpoints used may not be optimal for our purposes of converting from state inference predictions.

To evaluate potential output ranges for our SHAPE generation function, we reproduced experiments with native state directed NNTM (Section 3.1) and predicted state directed NNTM (Section 3.2) while varying the endpoints $a$ and $b$ of our SHAPE generation function, given in Equation~\ref{eq:shapegeneration}. As noted previously, a SHAPE value of $0.3603$ contributes no energy to the model; thus, it is only sensible to choose paired SHAPE values below $0.3603$, and unpaired SHAPE values above $0.3603$. NNTM software such as GTFold ignores negative SHAPE values, so paired nucleotides' generated SHAPE must lie between 0 and 0.3603. The results of this experiment are shown in Figure~\ref{heatmaps}.

In native state directed NNTM, increasing negative state SHAPE above 0.3603 and and decreasing positive state SHAPE below 0.3603 consistently increased performance. This is consistent with our expectations, as in this case we are increasing the energy of all base pairs involving nucleotides that remain unpaired in the native structure. Experiments with very large unpaired SHAPE values, such as $b = 20$, were similar to the largest values shown in Figure~\ref{heatmaps}, indicating that there is a ceiling of approximately 90\% test set accuracy for any method centered around SHAPE-directed NNTM such as ours.

The plot for predicted state directed NNTM shows a different picture, with increasing unpaired SHAPE values eventually leading to decreasing structure inference accuracy. That this pattern appears in the predicted state experiments but not native state experiments suggests that incorrectly assigning large SHAPE values to even a small number of natively paired nucleotides can be significantly harmful to NNTM performance. There is a large region of highest accuracy, with $a$ between 0 and 0.22 and $b$ between 0.7 and 1.5 giving accuracies near 68\%. The values of $a = 0.214$ and $b = 0.6624$ used in our results are near the boundary of this region. But we note that even optimal values of $a$ and $b$ give an accuracy of 69\%, only 2.5 percentage points above the experimentally motivated choices of $a$ and $b$ used in our results.

\subsection{Other RNA Types}

The dependence on large amounts of data inhibited our ability to extend this work to inference on other types of RNA, such as 5S and 23S ribosomal RNA, as we were unable to amass enough secondary structure data in these other contexts to successfully train a neural network. Instead, we explored applying our trained network to other RNA sequences, but found that a neural network trained on 16S RNA sequences produces much weaker results  on 5S and 23S rRNA sequences than on 16S sequences, see  Tables~\ref{Tab:testset5s}-\ref{Tab:testset23s}. We suspect this is primarily due to differences in sequence length. Tables~\ref{Tab:testset5s}-\ref{Tab:testset23s} also compare neural network state predictions with those from an HMM, and shows that the HMM is more capable of generalizing to these RNA families. However, we note that in all cases, HMM performance is poor relative to neural network performance on 16S sequences. This is consistent with our hypothesis that our neural network is recognizing long-range dependencies specific to the family of 16S sequences that HMMs are inherently unable to capture.


\begin{table*}[!ht]
\centering
\caption{State inference results on 5S sequences  \label{Tab:testset5s}}
\setlength{\tabcolsep}{4pt}
\begin{tabular}{|l|cc|cc|cc|}
\cline{1-7}
 & \multicolumn{2}{c|}{Accuracy} & \multicolumn{2}{c|}{PPV} & \multicolumn{2}{c|}{Sensitivity}\\
Sequence Name & LSTM & HMM & LSTM & HMM & LSTM & HMM \\
\cline{1-7}\
d.5.e.S.pombe.1.no  &  0.647  &  0.605  &  0.735  &  0.685  &  0.676  &  0.676\\
d.5.e.P.waltl.no  &  0.500  &  0.717  &  0.635  &  0.756  &  0.446  &  0.797\\
d.5.e.O.sativa.1.no  &  0.487  &  0.756  &  0.589  &  0.808  &  0.581  &  0.797\\
d.5.e.M.glyptostroboides.no  &  0.467  &  0.700  &  0.586  &  0.757  &  0.460  &  0.757\\
d.5.e.M.fossilis.no  &  0.442  &  0.675  &  0.571  &  0.733  &  0.378  &  0.743\\
d.5.b.M.luteus  &  0.516  &  0.650  &  0.700  &  0.709  &  0.449  &  0.782\\
d.5.b.M.genitalium  &  0.466  &  0.636  &  0.622  &  0.754  &  0.319  &  0.597\\
d.5.e.L.edodes  &  0.583  &  0.658  &  0.682  &  0.709  &  0.608  &  0.757\\
\cline{1-7}
Average  &  0.514  &  0.675  &  0.640  &  0.739  &  0.490  &  0.738\\
\cline{1-7}
\end{tabular}
\begin{flushleft}
Table of accuracy, PPV, and sensitivity for our LSTM-based state inference model and an order 4 HMM for 5S sequences. Average indicates the average metric for each sequence. 
\end{flushleft}
\end{table*}


\begin{table*}[!ht]
\centering
\caption{State inference results on 23S sequences  \label{Tab:testset23s}}
\setlength{\tabcolsep}{4pt}
\begin{tabular}{|l|cc|cc|cc|}
\cline{1-7}
 & \multicolumn{2}{c|}{Accuracy} & \multicolumn{2}{c|}{PPV} & \multicolumn{2}{c|}{Sensitivity}\\
Sequence Name & LSTM & HMM & LSTM & HMM & LSTM & HMM \\
\cline{1-7}
d.233.a.H.marismortui.1  &  0.640  &  0.682  &  0.647  &  0.709  &  0.840  &  0.771\\
d.233.a.T.celer  &  0.666  &  0.745  &  0.666  &  0.729  &  0.854  &  0.892\\
d.233.m.S.sinuspaulianus  &  0.541  &  0.632  &  0.407  &  0.478  &  0.678  &  0.543\\
PDB\_00784  &  0.652  &  0.669  &  0.714  &  0.764  &  0.803  &  0.732\\
PDB\_00616  &  0.602  &  0.655  &  0.595  &  0.655  &  0.829  &  0.767\\
PDB\_00953  &  0.595  &  0.663  &  0.567  &  0.622  &  0.868  &  0.856\\
PDB\_00503  &  0.612  &  0.658  &  0.610  &  0.668  &  0.830  &  0.760\\
PDB\_00846  &  0.642  &  0.664  &  0.709  &  0.763  &  0.793  &  0.725\\
PDB\_00776  &  0.552  &  0.602  &  0.487  &  0.524  &  0.860  &  0.770\\
\cline{1-7}
Average  &  0.611  &  0.663  &  0.600  &  0.657  &  0.817  &  0.757\\
\cline{1-7}
\end{tabular}
\begin{flushleft}
Table of accuracy, PPV, and sensitivity for our LSTM-based state inference model and an order 4 HMM for 23S sequences. Average indicates the average metric for each sequence. 
\end{flushleft}
\end{table*}


\section{Conclusion}

We introduced a method for secondary structure inference by connecting a deep learning method for state inference with previous work that leverages SHAPE-directed NNTM for structure inference. We tested this method on a set of 16S rRNA sequences with a wide range of MFE accuracies, and found large improvements over undirected MFE structures in most cases. These gains were not uniform throughout the test set, as several sequences with low undirected MFE accuracy experienced very little increase from predicted state direction. However, median increase in accuracy was more than 30 percentage points, and in the best case our method improved undirected NNTM accuracy by nearly 50 percentage points.

Experiments using sequences' native states to predict structure showed that a data-directed NNTM approach has the potential to improve MFE accuracy throughout the test set. However, it also uncovered significant limitations; regardless of the accuracy of the state inference method used to supply predictions, directed NNTM will not be able to produce high accuracy MFE structures for some sequences. These findings reinforce results from~\cite{Sukosd:2013rm} regarding the varying directability of sequences in the test set.

The performance of our state inference method was highly variable among test set sequences, with several clusters of high and low accuracy.
We found that high performance of our state inference method (and, in turn, the accuracy of secondary structures generated from these state predictions) was strongly linked to the similarity between a sequence's paired region distribution and that of the training set.

This finding highlights the connection between performance of our method and available secondary structure data. As with any application of machine learning, the state inference method presented here is only as good as the dataset used to train the model. The task of 16S rRNA state inference explored here is feasible primarily due to the particularly large number of known 16S rRNA secondary structures. We are hopeful that future increases in available data are able to increase the efficacy of this method, both on 16S rRNA sequences as well as on 5S and 23S rRNA.

\section*{Availability of data and materials}

The sequence and secondary structure data used to train the neural network and HMMs is available from the Comparative RNA Web (CRW) site repository, \url{https://doi.org/10.1186/1471-2105-3-2}. This data, along with the native secondary structure data for the test set, is available at \url{http://ms.uky.edu/~dwi239/rnastateinf-data.zip}. 

  
   



\section*{Acknowledgements}
We would like to thank John Hirdt, who had initially participated in this project, for many valuable discussions and suggestions.
DM thanks Christine Heitsch for introducing him into research on SHAPE-directed RNA structure prediction.   

\bibliographystyle{plain}
\bibliography{ref_rna.bib}      

\end{document}